\documentclass[aip,reprint]{revtex4-1}

\usepackage[utf8]{inputenc}
\usepackage[english]{babel}
\usepackage{graphicx}
\usepackage{amsmath}
\usepackage{amssymb}
\usepackage{hyperref}
\usepackage{xcolor}
\usepackage{bbold}
\usepackage{empheq}
\usepackage[most]{tcolorbox}

\newcommand{\e}{\varepsilon}
\def\w{\omega}

\newcommand{\ee}{\mathrm{e}}
\newcommand{\ii}{\mathrm{i}}
\newcommand{\dd}{\,\mathrm{d}}
\newcommand{\mO}{\mathcal{O}}

\newcommand{\mQ}{\mathcal{Q}}
\newcommand{\mF}{\mathcal{F}}

\newcommand{\bu}{{\bf u}}
\newcommand{\bx}{{\bf x}}
\newcommand{\bA}{{\bf A}}
\newcommand{\bJ}{{\bf J}}
\newcommand{\bM}{{\bf M}}
\newcommand{\bD}{{\bf D}}
\newcommand{\bG}{{\bf G}}
\newcommand{\bS}{{\bf S}}
\newcommand{\bX}{{\bf X}}
\newcommand{\bC}{{\bf C}}
\newcommand{\bH}{{\bf H}}

\begin{document}

\title{High-order phase reduction for coupled 2D oscillators}
\author{Erik T.K. Mau}
\email[]{erikmau@uni-potsdam.de}
\affiliation{Department of Physics and Astronomy, University of Potsdam, 
Karl-Liebknecht-Str. 24/25, D-14476 Potsdam-Golm, Germany}

\author{Michael Rosenblum}
\email[]{mros@uni-potsdam.de}
\affiliation{Department of Physics and Astronomy, University of Potsdam, 
Karl-Liebknecht-Str. 24/25, D-14476 Potsdam-Golm, Germany}

\author{Arkady Pikovsky}
\email[]{pikovsky@uni-potsdam.de}
\affiliation{Department of Physics and Astronomy, University of Potsdam,
Karl-Liebknecht-Str. 24/25, D-14476 Potsdam-Golm, Germany}

\date{\today}

\begin{abstract}
Phase reduction is a general approach to describe coupled oscillatory units in terms of their phases, assuming that the amplitudes are enslaved. The coupling should be small for such a reduction, but one also expects the reduction to be valid for finite coupling. This paper presents a general framework allowing us to obtain coupling terms in higher orders of the coupling parameter for generic two-dimensional oscillators and arbitrary coupling terms. The theory is illustrated with an accurate prediction of Arnold's tongue for the van der Pol oscillator exploiting higher-order phase reduction.
\end{abstract}

\maketitle

\begin{quotation}
The description of coupled oscillators is one of the basic problems in nonlinear dynamics. For weak coupling, the units remain oscillating but can adjust their phases. This adjustment results in synchronization and many other effects related to the phase dynamics. Representation  in terms of phases yields a simplified yet good quantitative characterization of the oscillating systems. To achieve this, one needs an accurate reduction from the original equations of motion to the phase dynamics equations, typically obtained approximately in the first order of the coupling strength. In this paper, we provide, for two-dimensional self-sustained oscillators, a theoretical perturbative framework for an improved reduction, which produces phase equations as expansions in the orders of a small parameter describing coupling.
\end{quotation}

\section{Introduction}

Phase approximation is a powerful tool widely used to analyze the dynamics of interacting self-sustained 
oscillators~\cite{winfree1980,Kuramoto-84,Hoppensteadt-Izhikevich-97,pikovsky2001,
Ermentrout-Terman-10,nakao2016phase,Monga_Wilson-Matchen-Moehlis-18,pietras2019network}.
This approach parametrizes each limit-cycle system with only one variable, the phase, and thus reduces the dimensionality of the problem. Behind this reduction lies the assumption that the amplitudes are enslaved variables following the evolution of the phases. In many cases, the reduced equations yield an analytical solution, with the celebrated Kuramoto model being an example. Even when one has to analyze the phase dynamics numerically, the approach greatly simplifies the original problem because only one variable has to be followed for each oscillator.  

Technically, the reduction to the phase dynamics relies on the smallness of the terms defining forcing or coupling of limit-cycle oscillators and, of course, on the proper definition of the phase. In the first order in the small parameter, one neglects the deviations of the amplitudes from the limit cycle, so only information about the phase in the vicinity of the limit cycle is needed (in the form of a set of isochrons or as a phase sensitivity function). However, one expects that the phase reduction is also valid for finite perturbation as long as the dynamics lie on an attracting high-dimensional torus spanned by the phases of interacting limit-cycle oscillators. For this, one needs to know the deviations of the amplitudes. 
Despite the number of attempts to account for these deviations and thus go beyond the first approximation in the coupling 
strength~\cite{kurebayashi2013,Monga_Wilson-Matchen-Moehlis-18,Wilson-Ermentrout-18,Mauroy-Mezic-18,wilson2019c,rosenblum2019, Rosenblum-Pikovsky-19a,Leon-Pazo-19,Perez-Seara-Huguet-20,gengel2021, kurebayashi2022,bick2023higherorder}, the high-order phase reduction remains challenging.  

In this communication, we describe the derivation of the high-order phase dynamics equations for generic two-dimensional limit-cycle oscillators. 
Our technique relies on the normal form 
phase-amplitude representation~\cite{Shilnikov_Shilnikov_Turaev_Chua-98} of the dynamics of a two-dimensional oscillator near the limit cycle. 
For an illustration of the normal form, consider the Stuart-Landau oscillator
\[
\dot a=(\zeta+\ii\eta)a-(1+\ii\alpha)|a|^2a\;,
\]
where $a$ is a complex variable and $\zeta,\eta,\alpha$ are parameters.
Writing $a=\rho e^{i\theta}$, one easily checks that for $\zeta>0$ the system has a stable circular limit cycle 
with radius $\sqrt{\zeta}$. The 
transformation~\cite{wilson2018a} 
\begin{align}
    \phi(\rho,\theta)=\theta-\ln(\rho/\sqrt{\zeta})\;,\quad r=c(1-\zeta/\rho^2)
    \label{eq:SL_transformation}
    \;,
\end{align}
where $c$ is any non-zero factor, 
recasts the systems to the autonomous normal form \cite{Shilnikov_Shilnikov_Turaev_Chua-98}
\begin{equation}
    \dot\phi=\eta-\alpha\zeta=\w\;,\quad \dot r=\kappa r
\label{eq:anf}    
\end{equation}
in the whole basin of the limit cycle.
Here $\phi$ is the phase, $\w$ is the frequency, and $\kappa=-2\zeta$ is the Floquet exponent. Variable $r$ quantifies deviation from the limit cycle; for brevity, we will call $r$ the amplitude (it is also known as the isostable variable). 

Essential for our analysis below is that for an {\em arbitrary smooth 2D system}, there exists a smooth variable substitution,  
reducing this system to the normal form \eqref{eq:anf} near a periodic trajectory, 
see Theorem 3.23 in Ref.~\onlinecite{Shilnikov_Shilnikov_Turaev_Chua-98}.  For higher-dimensional systems, the normal form can be more complex (e.g., for degenerate eigenvalues and in the case of resonances); this is a subject for future research.

In this communication, we exploit the perturbation technique to derive the phase coupling functions as a series in powers of the coupling strength $\e$. Our procedure is closely related to that of Gengel et al.~\cite{gengel2021} but is not restricted to the Stuart-Landau system.  First, we outline the derivation of the terms for a general system of $M$ coupled  2D units. Next, we explicitly write the terms up to the order $\e^3$ for two coupled oscillators and illustrate the approach by an application to the paradigmatic van der Pol model.  

\section{Generic many-body couplings}
\label{sec:generic_2D}

In this section, we sketch the derivation of the non-trivial $\mO(\e^2)$ terms in the phase reduction for $M$ generically coupled two-dimensional oscillators. Moreover, we outline the procedure to derive terms of arbitrary order.

We start by writing general equations for two-dimen\-sional limit-cycle systems with states $\bx_1, \dots, \bx_M \in \mathbb{R}^2$, indexed by $\mu \in \{1, \dots, M \}$:
\begin{align}
    \dot\bx_\mu &= \bG_\mu(\bx_\mu) + \e \bS_\mu(\bx_1, \dots, \bx_M)
    \label{eq:ODE_original}
    \,.
\end{align}
Here $\bG_\mu$ determines the autonomous evolution of oscillator $\mu$, and $\bS_\mu$ encodes the coupling of this unit to all other oscillators. We assume that $\bG_\mu$ and $\bS_\mu$ are sufficiently smooth functions in all arguments. Since all systems exhibit stable limit cycles, for each uncoupled unit there exists a smooth transformation to coordinates $\phi_\mu=\Phi_\mu(\bx_\mu)$ and $r_\mu=P_\mu(\bx_\mu)$ which obey linear normal form equations for each oscillator~\cite{Shilnikov_Shilnikov_Turaev_Chua-98} 
 (cf.~Eq.~\eqref{eq:anf}):
\begin{equation}
\dot\phi_\mu=\w_\mu\,,\quad \dot r_\mu=\kappa_\mu r_\mu\,,
\label{eq:pac}    
\end{equation}
where $\omega_\mu$ is the frequency of the limit-cycle oscillation and $\kappa_\mu<0$ is the real-valued Floquet exponent. 

The transformation functions fulfill the following equations
\begin{align}
    \w_\mu &= \nabla_{\bx_\mu} \Phi_\mu \cdot \bG_\mu \;,\\
    \kappa_\mu P_\mu &= \nabla_{\bx_\mu} P_\mu \cdot \bG_\mu
    \,.
\end{align}
Thus, the dynamics can be expressed in the phase-amplitude variables as
\begin{align}
    \dot \phi_\mu &= \w_\mu + \e \mQ_\mu(\Vec{\phi}, \Vec{r}) \;,
    \label{eq:ODE_phi} \\
    \dot r_\mu &= \kappa_\mu r_\mu + \e \mF_\mu( \Vec{\phi}, \Vec{r})\;,
    \label{eq:ODE_r}
\end{align}
where $\Vec{\phi} = (\phi_1, \dots, \phi_M)^\top$ and $\Vec{r} = (r_1, \dots, r_M)^\top$. Here, $\mQ_\mu$ and $\mF_\mu$ are the coupling functions in terms of the phases and the amplitudes: 
\begin{align}
    \mQ_\mu(\Vec{\phi}, \Vec{r})
    =& \nabla_{\bx_\mu} \Phi_\mu \cdot \bS_\mu(\bx_1, \dots, \bx_M) |_{\Vec{\phi}, \Vec{r}} \;,
    \label{eq:Q_def} \\
    \mF_\mu(\Vec{\phi}, \Vec{r})
    =& \nabla_{\bx_\mu} P_\mu \cdot \bS_\mu(\bx_1, \dots, \bx_M) |_{\Vec{\phi}, \Vec{r}}
    \,.
    \label{eq:F_def}
\end{align}
We remark that Eqs.~(\ref{eq:ODE_phi}-\ref{eq:F_def}) are equivalent to Eq.~\eqref{eq:ODE_original} as long as all states $\bx_\mu$ are in the domain of validity of transformations $\bx\to (\phi,r)$. One can argue that this domain extends to the whole basin of attraction of the respective limit cycle~\cite{DT}. However, we do not rely on this since the perturbation technique operates only in close vicinity of the cycle. So far, there has been no dimension reduction, and the new system has the same dimension $2M$. We also note that we use the normal form of all oscillators separately and do not perform the normal form analysis of the coupled system~\cite{ashwin2016hopf,nijholt2022emergent}; thus, no resonant/non-resonant conditions appear below.

We aim to derive a reduced model incorporating only the phases $\phi_\mu$. We achieve that by assuming that for a given (small) coupling strength $\e$, the dynamics, possibly after a transient time, is restricted to a $M$-dimensional torus fully parametrized by the phases. In other words, we assume that, in the long-time evolution, the amplitudes $r_\mu = R_\mu(\Vec{\phi})$ are functions of phases. Then, we write the asymptotic phase dynamics as 
\begin{align}
    \dot{\phi}_\mu = \w_\mu + \e \mQ_\mu(\Vec{\phi}, \Vec{R}(\Vec{\phi}))
    \label{eq:ODE_phi_reduced}
    \,,
\end{align}
where $\Vec{R} = (R_1, \dots, R_M)^\top$. Since for $\e=0$ one has $R_\mu(\Vec{\phi})=0$,  we expect that $R_\mu(\Vec{\phi})$ are small for small $\e$. Thus, we adopt a standard perturbation approach and represent the unknown functions as power series in $\e$:
\begin{align}
    \mQ_\mu(\Vec{\phi}) = \sum_{n=0}^\infty \mQ_{\mu;n}(\Vec{\phi}) \e^n\,,\; R_\mu(\Vec{\phi}) = \sum_{n=0}^\infty R_{\mu;n}(\Vec{\phi}) \e^n\,.
    \label{eq:Q_expansion_in_eps}
\end{align}
Although we expect that the expansion for $R_\mu$ starts with a linear term $\sim\e$, we start the formal expansion from $n=0$ for simplicity of notations; later, we will see that $R_{\mu;0}=0$. 

Keeping for definiteness the terms up to the order $\e^3$, we can represent the phase dynamics as:
\begin{align}
    \dot{\phi}_\mu = \w_\mu + \mQ_{\mu;0} \e + \mQ_{\mu;1} \e^2 + \mQ_{\mu;2} \e^3 + \mO(\e^4) 
    \,.
\end{align}
Here and in the following, we omit the functions' arguments $\Vec{\phi}$. 

To eliminate the amplitudes from the model, we need an equation that determines $R_\mu$; or, equivalently, a set of equations that determine $R_{\mu,n}$ in different orders $\e^n$. First, generally $\dot{r}$ is expressed as 
\begin{align}
    \dot{r}_\mu = \sum_{\nu=1}^M 
    \dot{\phi}_\nu \partial_{\phi_\nu} R_\mu 
    \label{eq:dot_r_with_R}
    \,.
\end{align}
Equating Eq.~\eqref{eq:dot_r_with_R} with the r.h.s of Eq.~\eqref{eq:ODE_r}, substituting $\dot{\phi}_\nu$ by the r.h.s of Eq.~\eqref{eq:ODE_phi} and rearranging terms yields
\begin{align}
    \kappa_\mu R_\mu - \sum_{\nu=1}^M \w_\nu \partial_{\phi_\nu} R_\mu 
    = \e \left( \sum_{\nu=1}^M \mQ_\nu \partial_{\phi_\nu} R_\mu - \mF_\mu \right)
    \label{eq:PDE_R}
    \,.
\end{align}
We remark that we use a notation $\mQ_\mu(\Vec{\phi}, \Vec{R}(\Vec{\phi}))=\mQ_\mu(\Vec{\phi})$ and analogously for $\mF_\mu$. Equation~\eqref{eq:PDE_R} with $2\pi$-periodic boundary conditions defines $R_\mu$, but is not immediately solvable. Therefore, we use the expansion \eqref{eq:Q_expansion_in_eps} intending to obtain a set of equations to solve for $R_{\mu;n}$ consecutively, starting with $n=1$. By inserting Eq.~\eqref{eq:Q_expansion_in_eps} for $R_\mu$, an $\e$-expansion for $\mF$ (analogous to $\mQ$ in Eq.~\eqref{eq:Q_expansion_in_eps}), and the term 
\begin{align}
    \mQ_\nu \partial_{\phi_\nu} R_\mu = \sum_{n=0}^\infty \sum_{\ell=0}^n \mQ_{\nu; \ell} \partial_{\phi_\nu} R_{\mu; n-\ell} \e^n
    \,,
\end{align}
which follows from the Cauchy product formula, into Eq.~\eqref{eq:PDE_R}, we obtain
\begin{align}
    \sum_{n=0}^\infty \left(\kappa_\mu R_{\mu;n} - \sum_{\nu=1}^M \w_\nu \partial_{\phi_\nu} R_{\mu;n} \right) \e^n
    = \e  \sum_{n=0}^\infty C_{\mu;n} \e^n
    \label{eq:PDE_expansion}
    \,.
\end{align}
Here, $C_{\mu;n}$ is defined as
\begin{align}
    C_{\mu; n} = \sum_{\nu=1}^M \sum_{\ell=0}^n \mQ_{\nu; \ell} \partial_{\phi_\nu} R_{\mu; n-\ell} - \mF_{\mu; n}
    \label{eq:Cn_definition}
    \,.
\end{align}

By matching terms of the same power in $\e$ in Eq.~\eqref{eq:PDE_expansion}, we obtain a set of equations determining all $R_{\mu;n}$. First, the terms of  $\mO(\e^0)$ yield $R_{\mu;0} = 0$, reflecting that the amplitudes vanish asymptotically without coupling. 
Next, by writing for clarity the arguments of the unknown terms explicitly, we obtain
\begin{align}
    \kappa_\mu R_{\mu;n}(\Vec{\phi}) - \sum_{\nu=1}^M \w_\nu \partial_{\phi_\nu} R_{\mu;n}(\Vec{\phi})
    = C_{\mu;n-1}(\Vec{\phi})
\end{align}
for all $ n \geq 1$. This is an inhomogeneous linear partial differential equation; the r.h.s. comes from the previous order of expansion and is a known function of $\Vec{\phi}$. Because the unknown functions $R_{\mu;n}(\Vec{\phi})$ are $2\pi$-periodic in their arguments, we straightforwardly write the Green's function of the equation in the Fourier space, cf.~\cite{gengel2021}. The solution reads:
\begin{align}
    R_{\mu;n} = \Xi_\mu[C_{\mu;n-1}]\;,
    \label{eq:R_n_solution}
\end{align}
where the operator $\Xi_\mu$ is defined as
\begin{align}
    \Xi_\mu[f](\Vec{\phi})
    &= \int_0^{2\pi} \frac{f(\Vec{\phi} - \Vec{\varphi})}{(2\pi)^M} \sum_{\Vec{\alpha} \in \mathbb{Z}^M} \frac{\ee^{-\ii \Vec{\alpha}\cdot \Vec{\varphi}}}{\kappa_\mu + \ii \Vec{\alpha} \cdot \Vec{\w}} \dd \Vec{\varphi}
    \label{eq:oper}
\end{align}
and $f$ is a $2\pi$-periodic test function. Here, $\Vec{\alpha} \cdot \Vec{\w} = \sum_{\nu=1}^M \alpha_\nu \w_\nu $ denotes the scalar product, and same for $ \Vec{\alpha}\cdot \Vec{\varphi}$. The operator $\Xi_\mu$ is linear and commutes with each $\partial_{\phi_\nu}$. 
Noteworthy, the denominator in \eqref{eq:oper} does not vanish for any $\Vec{\alpha}$; thus, there are no small divisors in this perturbation technique.

Equation~\eqref{eq:R_n_solution} yields an expression for each $R_{\mu;n}$. However, the functions $\mQ_{\nu;m}$ and $\mF_{\nu;m}$ appearing in $C_{\mu;n}$ are not directly available from the definitions of $\mQ$ and $\mF$ given by  Eqs.~(\ref{eq:Q_def}) and (\ref{eq:F_def}). To write an expression for $R_{\mu;n}$ in terms of the original coupling functions, we additionally need to  express $\mQ$ and $\mF$ from Eqs.~(\ref{eq:ODE_phi},\ref{eq:ODE_r}) as expansions in powers of $R_\mu$ (these expansions are well-defined because $R_\mu\sim \e$). We write
\begin{align}
    \mQ_\mu = \sum_{n=0}^\infty\;\; \sum_{\substack{\Vec{k} \text{ with } \sum_\nu k_\nu = n}}
    Q_{\mu; \Vec{k}} \prod_{\nu=1}^M R_\nu^{k_\nu}
    \label{eq:Q_expansion_in_r}
    \,,
\end{align}
where $\Vec{k} = (k_1, \dots, k_M) \in \mathbb{N}_0^M$ denotes a multi-index. This expression is practical, since $Q_{\mu; \Vec{k}}$ can be obtained from the derivative of $\mQ$ with respect to $r$ evaluated at the limit cycle:
\begin{align}
    Q_{\mu; \Vec{k}} = \left( \prod_{\nu=1}^M k_\nu! \partial_{r_\nu}^{k_\nu} \right)\mQ|_{\Vec{r}=0}
    \,.
\end{align}
We will use the same notation for  $\mF_\mu$:
\begin{align}
    \mF_\mu = \sum_{n=0}^\infty\;\; \sum_{\substack{\Vec{k} \text{ with } \sum_\nu k_\nu = n}}
    F_{\mu; \Vec{k}} \prod_{\nu=1}^M R_\nu^{k_\nu}\,.
    \label{eq:F_expansion_in_r}
\end{align}

By inserting the $\e$-expansions of $R_\mu$ from Eq.~\eqref{eq:Q_expansion_in_eps} into Eq.~\eqref{eq:Q_expansion_in_r} we identify each $\mQ_{\mu;n}$ (or $\mF_{\mu;n}$) with an expression containing  $R_{\nu;m}$ and $Q_{\nu; \Vec{k}}$ (or $F_{\nu; \Vec{k}}$) by collecting terms of the same power in $\e$. Substituting these expressions in \eqref{eq:Cn_definition}, we obtain the r.h.s. for determining the amplitudes in the next order, etc.
For the terms $\mO(1)$, we obtain
\begin{align}   
    \mQ_{\mu;0} = Q_{\mu;(0,\dots,0)}
    \label{eq:Q0}
\end{align}
directly. This corresponds to the standard Winfree form if the direction of the driving term $\bS_\mu$ is constant and independent of the state of the system $\mu$.

Next, the terms of $\mO(\e)$ yield
\begin{align}
    \mQ_{\mu;1} = \sum_{\nu=1}^M Q_{\mu; \Vec{e}_\nu} R_{\nu;1}
    \,,
\end{align}
where $\Vec{e}_\nu$ is $0$ everywhere except for the $\nu$-th place, where it is $1$.
We obtain $C_{\mu;0}$ from Eq.~\eqref{eq:Cn_definition} as
\begin{align}
     C_{\mu;0} &= - \mF_{\mu;0} = - F_{\mu;(0,\dots, 0)}
     \label{eq:C0}
\end{align}
and write according to Eq.~\ref{eq:R_n_solution}:
\begin{align}
    R_{\mu;1} = -\Xi_\mu[F_{\mu;(0,\dots, 0)}] \;.
    \label{eq:R1}
\end{align}
We finally obtain
\begin{align}   
    \mQ_{\mu;1} = - \sum_{\nu=1}^M Q_{\mu; \Vec{e}_\nu} \Xi_\nu[F_{\nu;(0, \dots, 0)}] 
    \label{eq:Q1}
    \,.
\end{align}
Eq.~\eqref{eq:Q1} yields the first non-trivial term of the phase-reduced model of generically coupled two-dimensional oscillators. We demonstrate the advantage of the corresponding $\mO(\e^2)$ phase model over the $\mO(\e)$ model in Sec.~\ref{sec:example_driven_system}. 
We remind that we can analogously conclude
\begin{align}
    \mF_{\mu;1} = - \sum_{\nu=1}^M F_{\mu; \Vec{e}_\nu} \Xi_\nu[F_{\nu;(0, \dots, 0)}]
    \,.
\end{align}

Now, we highlight that the procedure used to derive $\mQ_{\mu;1}$ and $\mF_{\mu;1}$ can be further exploited to derive $\mQ_{\mu;n}$ and $\mF_{\mu;n}$ by iterations for an arbitrarily large $n$. Assume we have appropriate expressions for all functions up to and including order $n$, i.e., $\mQ_{\mu;n}$ and $\mF_{\mu;n}$ as well as $R_{\mu;n}$. We want to obtain the terms of next highest order $\mQ_{\mu;n+1}$ and $\mF_{\mu;n+1}$ and $R_{\mu;n+1}$. To start with the latter, we use Eq.~\eqref{eq:R_n_solution}, which requires $C_{\mu;n}$. We check in Eq.~\eqref{eq:Cn_definition} that $C_{\mu;n}$ requires only terms up to order $n$. Thus, we obtain $R_{\mu;n+1}$. To infer $\mQ_{\mu;n+1}$ and $\mF_{\mu;n+1}$, we need to collect the terms of $\mO(\e^{n+1})$ in Eq.~\eqref{eq:Q_expansion_in_r}. Those will contain the accessible functions $Q_{\nu;\Vec{k}}$ and $F_{\nu;\Vec{k}}$ and $R_{\nu;m}$ where $m \in \{1, \dots, n+1\}$ and $\nu \in \{1,\dots, M\}$. Thus, we also obtain $\mQ_{\mu;n+1}$ and $\mF_{\mu;n+1}$, what closes the iteration loop.

Though the evaluation of $\mQ_{\mu; n}$ and $\mF_{\mu; n}$ becomes cumbersome very quickly, one can, in principle, continue to derive them for an arbitrarily large $n$ by repeating that procedure. The required functions $Q_{\nu;\Vec{k}}$ and $F_{\nu;\Vec{k}}$ can be computed from the phase-amplitude transformation in the vicinity of the limit cycle, which can be obtained numerically (see Ref.~\onlinecite{wilson2020a} and appendix \ref{apx:numerics_reponse_curves}).
We will demonstrate this procedure by deriving $\mQ_{\mu;2}$, and consequently the $\mO(\e^3)$ phase model, for the special case of two coupled oscillators in Sec.~\ref{sec:two_oscillators}.

\section{Higher-order coupling functions for two coupled oscillators}
\label{sec:two_oscillators}
For the case of two coupled oscillators, many expressions simplify.
The next term in the expansion for $\mQ$ reads
\begin{equation}
\begin{gathered}
     \mQ_{\mu;2} =  
    Q_{\mu;(2,0)} R^2_{1;1} + 
    Q_{\mu;(1,0)} R_{1;2}  \\ 
   + Q_{\mu;(1,1)} R_{1;1}R_{2;1}  +
    Q_{\mu;(0,1)} R_{2;2} +
    Q_{\mu;(0,2)} R^2_{2;1}\,.
    \end{gathered}
    \label{eq:Q2_preliminary}
\end{equation}
To evaluate that, we require $R_{\mu;2}$ and, thus, also $C_{\mu;1}$. Setting $n=1$ in Eq.~\eqref{eq:Cn_definition}, we obtain 
\begin{align}
    C_{\mu;1} &= \mQ_{1; 0} \partial_{\phi_1} R_{\mu; 1} + \mQ_{2; 0} \partial_{\phi_2} R_{\mu; 1} - \mF_{\mu; 1}\;.
\end{align}
Replacing $R_{\mu; 1}$ (Eq.~\eqref{eq:R1}),  $\mQ_{\mu; 0}$ (Eq.~\eqref{eq:Q0}), and $\mF_{\mu; 1}$ (analogous to Eq.~\eqref{eq:Q1}) yields
\begin{align}
    C_{\mu;1}
    = \quad &  
    Q_{1;(0,0)} \Xi_\mu [\partial_{\phi_1} F_{\mu;(0,0)}] + 
    Q_{2;(0,0)} \Xi_\mu [\partial_{\phi_2} F_{\mu;(0,0)}]
    \nonumber \\ -& 
    F_{\mu;(1,0)} \Xi_1[F_{1;(0,0)}] - F_{\mu;(0,1)} \Xi_2[F_{2;(0,0)}]
    \,.
\end{align}
Thus, we get
\begin{align}
    R_{\mu;2} = \quad &  
    \Xi_\mu[Q_{1;(0,0)} \Xi_\mu [\partial_{\phi_1} F_{\mu;(0,0)}]]
    \nonumber \\ +& 
    \Xi_\mu[Q_{2;(0,0)} \Xi_\mu [\partial_{\phi_2} F_{\mu;(0,0)}]]
    \nonumber \\ -& 
    \Xi_\mu[F_{\mu;(1,0)} \Xi_1[F_{1;(0,0)}]]
    \nonumber \\ -&
    \Xi_\mu[F_{\mu;(0,1)} \Xi_2[F_{2;(0,0)}]]
    \label{eq:R2}
    \,.
\end{align}

We obtain $\mQ_{\mu;2}$ by inserting $R_{\mu;1}$ (Eq.~\eqref{eq:R1}) and $R_{\mu;2}$ (Eq.~\eqref{eq:R2}) into Eq.~\eqref{eq:Q2_preliminary} as
\begin{align}
    \mQ_{\mu;2} = \quad &  
    Q_{\mu;(2,0)} (\Xi_1[F_{1;(0,0)}])^2
    \nonumber \\ +& 
    Q_{\mu;(1,0)} \Xi_1[Q_{1;(0,0)} \Xi_1[\partial_{\phi_1} F_{1;(0,0)}]]
    \nonumber \\ +&
    Q_{\mu;(1,0)} \Xi_1[Q_{2;(0,0)} \Xi_1[\partial_{\phi_2} F_{1;(0,0)}]]
    \nonumber \\ -&
    Q_{\mu;(1,0)} \Xi_1[F_{1;(1,0)} \Xi_1[F_{1;(0,0)}]] 
    \nonumber \\ -&
    Q_{\mu;(1,0)} \Xi_1[F_{1;(0,1)} \Xi_2[F_{2;(0,0)}]]
    \nonumber \\ +& 
    Q_{\mu;(1,1)} \Xi_1[F_{1;(0,0)}]\Xi_2[F_{2;(0,0)}]
    \nonumber \\ +& 
    Q_{\mu;(0,1)} \Xi_2[Q_{1;(0,0)} \Xi_2[\partial_{\phi_1} F_{2;(0,0)}]]
    \nonumber \\ +&
    Q_{\mu;(0,1)} \Xi_2[Q_{2;(0,0)} \Xi_2[\partial_{\phi_2} F_{2;(0,0)}]]
    \nonumber \\ -&
    Q_{\mu;(0,1)} \Xi_2[F_{2;(1,0)} \Xi_1[F_{1;(0,0)}]]
    \nonumber \\ -&
    Q_{\mu;(0,1)} \Xi_2[F_{2;(0,1)} \Xi_2[F_{2;(0,0)}]]
    \nonumber \\ +& 
    Q_{\mu;(0,2)} (\Xi_2[F_{2;(0,0)}])^2
    \,.
    \label{eq:Q2}
\end{align}
This completes the derivation of the coupling function for two oscillators up to order $\e^3$.

\section{Higher-order coupling for a driven system}
\label{sec:example_driven_system}

We illustrate the general results of Sec.~\ref{sec:generic_2D} for the simplest case of a harmonically driven van der Pol oscillator
\begin{align}
    \dot x &= y + \e \cos(\phi_2)\;, \label{eq:van_der_Pol_driven1}\\
    \dot y &= a y(1-x^2) -x\;, \label{eq:van_der_Pol_driven2}\\
    \dot \phi_2 &= \w_2\;,
    \label{eq:van_der_Pol_driven3}
\end{align}
where we set $a=1.4$. In the following, we will refer to it as the 'full model'. Here, $\bx_1 = (x,y)$ is the state of the van der Pol oscillator, i.e., oscillator $1$. Oscillator $2$ represents a mere driving: since $\bS_2=0$ here, the response functions $\mQ_2=\mF_2=0$ and the amplitude deviation $r_2=0$ of the second oscillator vanish.

Since $\bS_1$ is independent of the state of oscillator $1$ and constant in direction (because it enters only in one Eq.~\eqref{eq:van_der_Pol_driven1}), we can factorize the
coupling functions: $\mQ_1(\phi_1, r_1, \phi_2) = Z(\phi_1, r_1) \cos(\phi_2)$, where $Z = \nabla \Phi \cdot (1,0)^\top$, and $\mF_1(\phi_1, r_1, \phi_2) = I(\phi_1, r_1) \cos(\phi_2)$, where $I = \nabla P \cdot (1,0)^\top$. Evaluated at the limit cycle $r_1=0$, $Z(\phi_1, 0)$ and $I(\phi_1, 0)$ represent the standard phase and amplitude response curves, and we denote them as $Z_0(\phi_1)$ and $I_0(\phi_1)$ in the following. Moreover, we define $Z_1(\phi_1) = \frac{\partial}{\partial r_1} \left. Z(\phi_1, r_1)\right|_{r_1=0}$. In Appendix~\ref{apx:numerics_reponse_curves}, we provide details on how the system-specific functions $Z_0, Z_1, I_0$ are determined numerically.

Thus, the derivatives of the response functions $\mQ_1$, $\mF_1$ with respect to $r_1$ which are necessary for the $\mO(\e^2)$ model read
\begin{align}
    Q_{1,(0,0)}(\phi_1, \phi_2) &= Z_0(\phi_1)\cos(\phi_2)\;, \\
    Q_{1,(1,0)}(\phi_1, \phi_2) &= Z_1(\phi_1)\cos(\phi_2)\;, \\
    F_{1,(0,0)}(\phi_1, \phi_2) &= I_0(\phi_1)\cos(\phi_2)\;,
\end{align}
and we conclude
\begin{align}
    \mQ_{1;0}(\phi_1, \phi_2) &= Z_0(\phi_1)\cos(\phi_2)\;,\\
    \mQ_{1;1}(\phi_1, \phi_2) &= -Z_1(\phi_1)\cos(\phi_2) \Xi_1[I_0(\phi_1)\cos(\phi_2)]
    \,.
\end{align}
Operator $\Xi_1$ is evaluated using a finite number of $17$ Fourier modes to approximate $I_0$. Since $\Xi_\mu$ resembles a convolution, the evaluation in the Fourier space is essentially a product of the Fourier modes. We now construct the $\mO(\e^2)$ model
\begin{align}
    \dot{\phi}_1 &= \w_1 + \mQ_{1;0}(\phi_1, \phi_2)\e + \mQ_{1;1}(\phi_1,\phi_2) \e^2 \;,\\
    \dot{\phi}_2 &= \w_2
    \;,
\end{align}
where $\w_1$ and $\kappa_1$ (required to evaluate the operator $\Xi_1$) are determined by the autonomous ($\e=0$) periodic solution of the van der Pol oscillator. The coupling strength $\e$ and the driving frequency $\w_2$ are free parameters.

\begin{figure}[!ht]
    \includegraphics[width=0.48\textwidth]{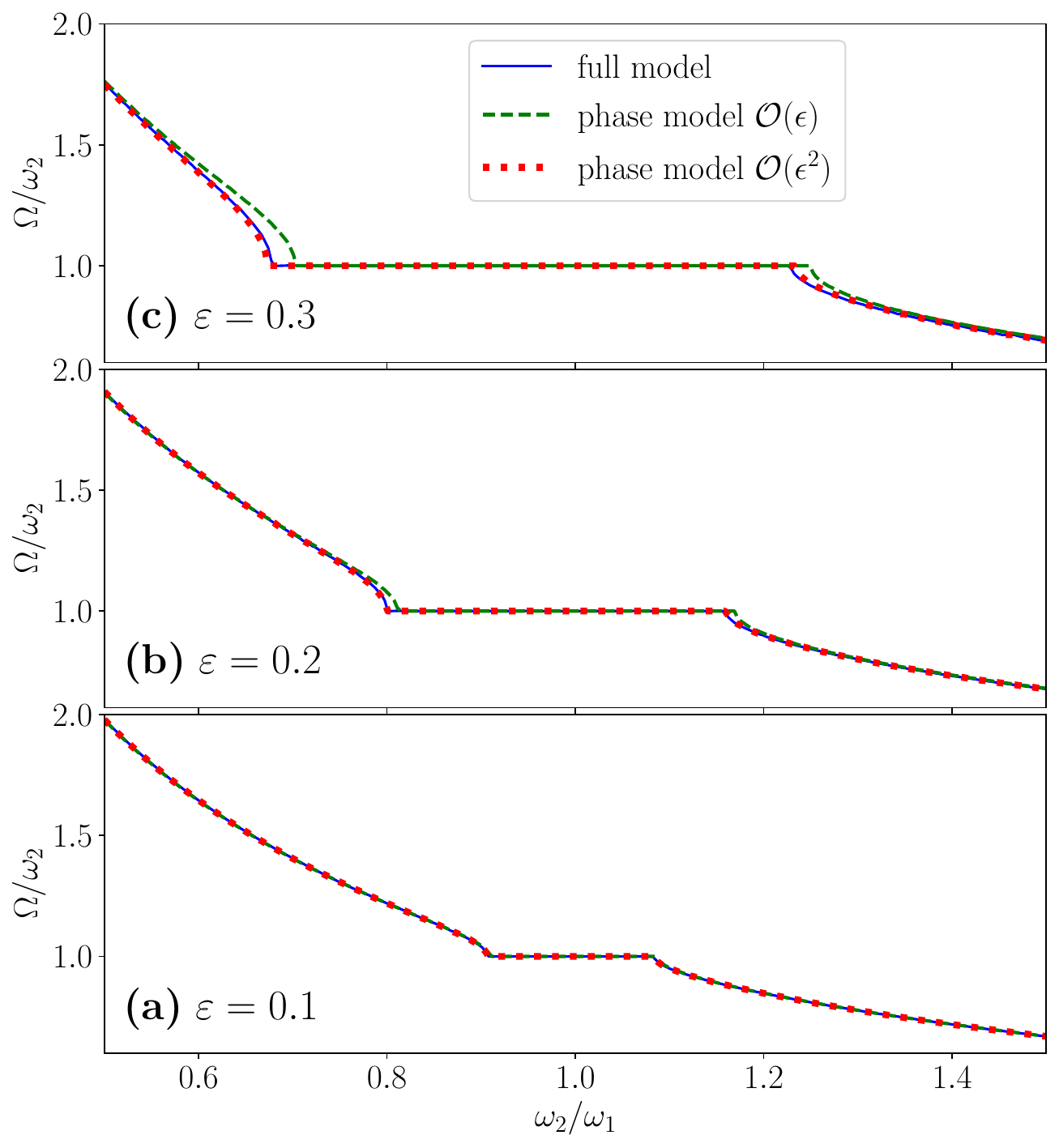}
    \caption{The ratio of the observed frequency $\Omega$ (see text) and driving frequency $\w_2$ vs $\w_2/\w_1$ for $\e=0.1$ (a), $\e=0.2$ (b) and $\e=0.3$ (c). We show the results for the full model (blue solid line), the phase models of order $\e$ (green dashed line), and $\e^2$ (red dotted line).
    All models show the $1:1$ synchronization region.
    For $\e=0.1$ (a), all the curves almost coincide. While the borders of the $\mO(\e)$ model become shifted relative to the full model as $\e$ grows, the $\mO(\e^2)$ phase model matches the full model almost perfectly even at the strongest coupling $\e=0.3$.}
    \label{fig:vdP_sync}
\end{figure}

In the following, we compare the $\mO(\e)$ and $\mO(\e^2)$ models against the full model by determining the borders of the Arnold tongue for a fixed $\e$ numerically. For this, we vary $\w_2$, integrate the full model and both phase models, and compute their respective observed frequencies by 
$\Omega = |\varphi(t_0+\tau)-\varphi(t_0)|/\tau$. 
For the phase models, $\varphi$ is an unwrapped phase $\varphi=\phi_1$, and for the full model $\varphi$ is unwrapped $\arctan(y/x)$.~\footnote{We fix $\tau= 2000$ and $t_0 = 2000$. The initial states for the phase models are $(\phi_1, \phi_2) = (0,0)$ and the initial state for the full model is $(x,y,\phi_2) = (1,0,0)$.} Fig.~\ref{fig:vdP_sync} demonstrates that the derived $\mO(\e^2)$ phase model reproduces the effective frequency of the full model more accurately than the $\mO(\e)$ model, as $\e$ becomes larger.

For another test, we employed the driven SL model. Here, all characteristics of the oscillator, such as $\w_1$, $\kappa_1$ as well as $Z_0, Z_1, I_0$ can be obtained analytically since the transformations to normal form (phase-isostable) coordinates~\ref{eq:SL_transformation} are known. We chose a driving that contains a first and a second harmonic term. In that way, the $\mO(\e^2)$ phase model predicts the appearance of the $1:2$ and $3:2$ synchronization regions that are not present in the $\mO(\e)$ phase model. Numerics demonstrates a good correspondence of these Arnold's tongues to those in the full model (not shown). Thus, higher-order corrections not only increase the accuracy of the predictions but can lead to novel features not captured by the leading order, cf.~\cite{PhysRevE.104.054202}.

\section{Discussion}

Summarizing, we presented a general framework for performing phase reduction for limit-cycle oscillators in higher orders of a small parameter that determines the coupling and/or forcing. The approach exploits the normal-form coordinates introduced separately for each oscillator. According to the general theory of smooth dynamical systems, these coordinates exist for arbitrary two-dimensional oscillators with a limit cycle. The situation is more subtle in a higher-dimensional case and will be considered elsewhere; see also~\cite{vondergracht2023parametrisation}. While the normal coordinates are proven to exist, their practical implementation needs a strongly nonlinear analysis of the original equations, which can be performed numerically, as outlined in appendix~\ref{apx:numerics_reponse_curves}. The resulting coupling terms (Eqs.~(\ref{eq:Q0}, \ref{eq:Q1}) in general $M$-dimensional case and Eq.~\eqref{eq:Q2} for two coupled oscillators) are obtained through an iterative procedure, the only nontrivial element of which is solving a linear PDE for the amplitude deviations.

We stress that our approach applies to a generic coupling, not only to a pair-wise one, as often assumed in the analysis of oscillator populations. Of course, for pair-wise couplings, some expressions can be potentially simplified. We notice that for a pair-wise coupling, the 
phase-coupling terms remain pair-wise in the leading order, but higher-order terms contain many-body (triplet, quadruplet, etc.) interactions; see discussion in Ref.~\onlinecite{gengel2021}. We also stress that the present approach does not allow for calculating the range of validity (in terms of the perturbation strength $\e$) of the derived phase equations.
One can assume that the equations are valid as long as the amplitudes are algebraic functions of the phases. This is equivalent to the condition that an invariant torus exists in the system's phase space. This condition is also used in a similar technique to obtain high-order phase equations~\cite{vondergracht2023parametrisation}, which appeared after the present study had been completed.

\begin{acknowledgments}
E.T.K.M. acknowledges financial support from Deutsche Forschungsgemeinschaft (DFG, German Research Foundation), Project-ID 424778381 – TRR 295. We are thankful to D. Turaev for valuable discussions.
\end{acknowledgments}

\section*{Data availability}
All numerical experiments are described in the paper. Computer
codes can be obtained from the corresponding author upon reasonable request.

\appendix

\section{Obtaining the normal form (phase-isostable) transformation close to the limit cycle}
\label{apx:numerics_reponse_curves}

To compute the phase-amplitude coupling functions  $\mQ$ and $\mF$ that are necessary for the construction of a phase model, we need the phase-amplitude transformations $\Phi$ and $P$, at least in the vicinity of the limit cycle. These functions allow for representing the corresponding Jacobian matrix ${\bf A}$ of the transformation as an expansion in powers of $r$:
\begin{align}
    {\bf A} = \begin{pmatrix} 
    \partial_x \Phi & \partial_y \Phi \\
    \partial_x P & \partial_y P \\    
    \end{pmatrix}
    = \sum_{n=0}^\infty {\bf A}_n \frac{r^n}{n!}
    \,.
\end{align}
With that construction, the information for the $n$-th derivative of $\mQ$ and $\mF$ with respect to $r$ evaluated at $r=0$ is contained in the phase-dependent matrices $\bA_1, \dots, \bA_n$. For the purpose of deriving an $\mO(\e^2)$ phase model we need $\bA_0$ and $\bA_1$. We will detail their inference from the dynamical model in the following.

First, we define the reverse transformation from the phase and the amplitude 
to Cartesian coordinates $\bx = (x,y)^\top$ as $\bX(\phi,r) = (X(\phi, r),Y(\phi, r))^\top$ and write $\bX$ as an asymptotic expansion in $r$:
\begin{align}
    \bX(\phi,r) &= \sum_{n=0}^\infty \bX_n(\phi) \frac{r^n}{n!}\;,
    \label{eq:x_expansion_in_psi}
\end{align}
where $\bX_n = (X_n,Y_n)^\top$. Its Jacobian matrix $\bJ$ reads
\begin{align}
    {\bf J} = \begin{pmatrix} 
    \partial_\phi X & \partial_r X \\
    \partial_\phi Y & \partial_r Y \\
    \end{pmatrix}
    = \sum_{n=0}^\infty {\bf J}_n \frac{r^n}{n!}\;,
\end{align}
where
\begin{align}
    {\bf J}_n = \begin{pmatrix} 
    \partial_\phi X_n & X_{n+1} \\
    \partial_\phi Y_n & Y_{n+1} \\    
    \end{pmatrix}
    \,.
\end{align}
Since $\bJ$ is the inverse of $\bA$, the equation $\mathbb{1} = {\bf A}{\bf J}$ holds and we conclude 
$
    \bA_0 = \bJ_0^{-1}
$ by setting $r=0$. Moreover, we obtain
$
   \partial_r \bA = - \bA \cdot \partial_r \bJ \cdot \bA
$
by differentiating with respect to $r$, ultimately leading to 
$
    \bA_1 = - \bA_0 \bJ_1 \bA_0
    \,
$
by setting $r=0$. Thus, to obtain the matrix elements of $\bA_1$ and $\bA_0$, we need $\bJ_0$ and $\bJ_1$, hence $\bX_0$, $\bX_1$ and $\bX_2$.

The limit cycle $\bX_0$  can be obtained by integrating the system forward in time sufficiently long. This also yields the period $T=2\pi/\omega$ of the system. An arbitrarily chosen point on the limit cycle $\bX_0(0)$ is assigned the phase $\phi=0$.

In the next step, we compute $\bX_1$ and $\kappa$ from the linearization around the limit cycle. Using the autonomous phase-amplitude dynamical equations
$
    \dot \phi = \omega $, $    \dot r = \kappa $
 we find 
\begin{align}
    \dot\bx  = \bG(\bX(\phi,r)) = \sum_{n=0}^\infty (\w \partial_\phi \bX_n + n\kappa \bX_n) \frac{r^n}{n!}
    \label{eq:G_expansion_in_r}
\end{align}
by differentiating Eq.~\eqref{eq:x_expansion_in_psi} with respect to time $t$. By evaluating $\partial_r \bG(\bX(\phi,r))$ at $r=0$ we thus conclude
\begin{align}
   \bD_\bG(\bX_0(\phi)) \cdot \bX_1 = \w \partial_\phi \bX_1 + \kappa \bX_1\;,
\end{align}
where $\bD_\bG$ is the Jacobian of $\bG$ defined by
\begin{align}
    \bD_\bG = 
    \begin{pmatrix} 
    \partial_x \bG_x & \partial_y \bG_x \\
    \partial_x \bG_y & \partial_y \bG_y \\
    \end{pmatrix}
    \,.
\end{align}
Rearranging terms, we find
$
    \w \partial_\phi \bX_1 =  (\bD_\bG(\bX_0(\phi))-\kappa \mathbb{1}) \cdot \bX_1
$
and given the transformation
\begin{align}    
\bu_1(t) = \ee^{\kappa t} \bX_1(\omega t)
    \label{eq:u1_def}
\end{align}
we arrive at the standard linearized dynamical equation for small deviations around the limit cycle
\begin{align}
    \dot{\bu}_1 = \bD_\bG(\bX_0(\omega t)) \cdot \bu_1
    \label{eq:linearized_deviation_limit_cycle}
    \,.
\end{align}
According to Floquet theory, Eq.~\eqref{eq:linearized_deviation_limit_cycle} is solved by
\begin{align}
    \bu_1(t) = \bM(t) \cdot \bu_1(0)\;,
    \label{eq:u1_solution}
\end{align}
where $\bM$ is the principal fundamental solution, that we find by numerically integrating Eq.~\eqref{eq:linearized_deviation_limit_cycle} in the interval $0 \leq t \leq T$ with initial conditions $\bu_1(0) \in \{(1,0)^\top, (0,1)^\top\}$. The non-unity eigenvalue of the monodromy matrix $\bM(T)$ is the Floquet multiplier $\ee^{\kappa T}$ from which we derive the real Floquet exponent $\kappa < 0$.

Thus, transforming back to $\bX_1$ yields 
$
    \bX_1(\phi) = \ee^{-\kappa \phi/\omega} \bM\left(\phi/\omega\right) \bX_1(0)$, $
    \quad \phi \in [0,2\pi)$.
Since we require $\bX_1(2\pi) = \bX_1(0)$, the initial condition $\bX_1(0)$ has to be an eigenvector of $\bM(T)$ corresponding to the non-unity Floquet multiplier. We fix the scaling of the isostable coordinate by choosing $||\bX_1(0)||=-1$ inside the limit cycle, where $||\cdot||$ denotes the standard Euclidean norm.

To find $\bX_2$, we again employ Eq.~\eqref{eq:G_expansion_in_r} to obtain
\begin{align}
   \bC(\bX_0, \bX_1) + \bD_\bG(\bX_0) \cdot \bX_2 = \w \partial_\phi \bX_2 + 2\kappa \bX_2
   \label{eq:ODE_X2}
   \,.
\end{align}
Here and in the following, we omit the notation of argument $\phi$ for conciseness. The term $\bC$ is defined component-wisely as
\begin{align}
    \bC_{x,y}(\bX_0, \bX_1) &= \bX_1^\top \cdot \bH_{\bG_{x,y}}(\bX_0) \cdot \bX_1 
    \label{eq:Cy_def}
\end{align}
where $\bH$ is the Hessian matrix defined as
\begin{align}
    \bH_f = 
    \begin{pmatrix} 
    \partial^2_x f & \partial_y \partial_x f \\
    \partial_x \partial_y f & \partial^2_y f \\
    \end{pmatrix}
    \,.
\end{align}
We get the equation for $\bX_2$ by rearranging terms as
\begin{align}
   \w \partial_\phi \bX_2 = (\bD_\bG(\bX_0) - 2\kappa \mathbb{1})\bX_2  + \bC(\bX_0, \bX_1)
   \,.
\end{align}
By introducing $\bu_2$ as
$
    \bu_2(t) = \ee^{2\kappa t} \bX_2(\w t)\;,
$
we derive its dynamical equation as 
\begin{align}
    \dot{\bu}_2 = \bD_\bG(\bX_0(\w t)) \bu_2 + \bC(\bX_0(\w t), \bu_1(t))\;,
    \label{eq:ODE_u2}
\end{align}
where we used Eq.~\eqref{eq:u1_def} to replace $\bX_1$ by $\bu_1$. Since $\bX_0$ and $\bu_1$ are known, this inhomogeneous linear ODE can be solved with Floquet theory. In fact, we employ the principal fundamental solution $\bM(t)$ from Eq.~\eqref{eq:u1_solution} to write the general solution as
\begin{align}
    \bu_2(t) = \bM(t) \bu_2(0) + \Tilde{\bu}_2(t)
    \,.
\end{align}
Here, $\Tilde{\bu}_2$ is the special solution to Eq.~\eqref{eq:ODE_u2} with initial condition $\Tilde{\bu}_2=0$. Thus, we obtain
\begin{align}
    \bX_2(\phi) =  \ee^{-2\kappa \phi/\w} \left( \bM(\phi/\w) \bX_2(0) + \Tilde{\bu}_2(\phi/\w) \right)\;,
\end{align}
where the initial condition $\bX_2(0)=\bX_2(2\pi)$ has to satisfy
\begin{align}
    \bX_2(0) = (\mathbb{1}-\ee^{-2\kappa T} \bM(T))^{-1}\ee^{-2\kappa T} \Tilde{\bu}_2(T)
    \,
\end{align}
to ensure the $2\pi$-periodicity. With $\bX_0$, $\bX_1$ and $\bX_2$, we construct $\bJ_0$ and $\bJ_1$, and thus compute $\bA_0$ and $\bA_1$.

For the case of a non-parametrically driven oscillator with the driving term acting in $x$-direction, as presented in Sec.~\ref{sec:example_driven_system}, the response functions follow directly as $Z_0 = (A_0)_{11}$, $Z_1 = (A_1)_{11}$ and $I_0 = (A_0)_{21}$.

\nocite{*}


\begin{thebibliography}{28}%
\makeatletter
\providecommand \@ifxundefined [1]{%
 \@ifx{#1\undefined}
}%
\providecommand \@ifnum [1]{%
 \ifnum #1\expandafter \@firstoftwo
 \else \expandafter \@secondoftwo
 \fi
}%
\providecommand \@ifx [1]{%
 \ifx #1\expandafter \@firstoftwo
 \else \expandafter \@secondoftwo
 \fi
}%
\providecommand \natexlab [1]{#1}%
\providecommand \enquote  [1]{``#1''}%
\providecommand \bibnamefont  [1]{#1}%
\providecommand \bibfnamefont [1]{#1}%
\providecommand \citenamefont [1]{#1}%
\providecommand \href@noop [0]{\@secondoftwo}%
\providecommand \href [0]{\begingroup \@sanitize@url \@href}%
\providecommand \@href[1]{\@@startlink{#1}\@@href}%
\providecommand \@@href[1]{\endgroup#1\@@endlink}%
\providecommand \@sanitize@url [0]{\catcode `\\12\catcode `\$12\catcode
  `\&12\catcode `\#12\catcode `\^12\catcode `\_12\catcode `\%12\relax}%
\providecommand \@@startlink[1]{}%
\providecommand \@@endlink[0]{}%
\providecommand \url  [0]{\begingroup\@sanitize@url \@url }%
\providecommand \@url [1]{\endgroup\@href {#1}{\urlprefix }}%
\providecommand \urlprefix  [0]{URL }%
\providecommand \Eprint [0]{\href }%
\providecommand \doibase [0]{http://dx.doi.org/}%
\providecommand \selectlanguage [0]{\@gobble}%
\providecommand \bibinfo  [0]{\@secondoftwo}%
\providecommand \bibfield  [0]{\@secondoftwo}%
\providecommand \translation [1]{[#1]}%
\providecommand \BibitemOpen [0]{}%
\providecommand \bibitemStop [0]{}%
\providecommand \bibitemNoStop [0]{.\EOS\space}%
\providecommand \EOS [0]{\spacefactor3000\relax}%
\providecommand \BibitemShut  [1]{\csname bibitem#1\endcsname}%
\let\auto@bib@innerbib\@empty
\bibitem [{\citenamefont {Winfree}(1980)}]{winfree1980}%
  \BibitemOpen
  \bibfield  {author} {\bibinfo {author} {\bibfnamefont {A.~T.}\ \bibnamefont
  {Winfree}},\ }\href {\doibase 10.1007/978-3-662-22492-2} {\emph {\bibinfo
  {title} {The {Geometry} of {Biological} {Time}}}}\ (\bibinfo  {publisher}
  {Springer Berlin Heidelberg},\ \bibinfo {address} {Berlin, Heidelberg},\
  \bibinfo {year} {1980})\BibitemShut {NoStop}%
\bibitem [{\citenamefont {Kuramoto}(1984)}]{Kuramoto-84}%
  \BibitemOpen
  \bibfield  {author} {\bibinfo {author} {\bibfnamefont {Y.}~\bibnamefont
  {Kuramoto}},\ }\href@noop {} {\emph {\bibinfo {title} {Chemical Oscillations,
  Waves and Turbulence}}}\ (\bibinfo  {publisher} {Springer},\ \bibinfo
  {address} {Berlin},\ \bibinfo {year} {1984})\BibitemShut {NoStop}%
\bibitem [{\citenamefont {Hoppensteadt}\ and\ \citenamefont
  {Izhikevich}(1980)}]{Hoppensteadt-Izhikevich-97}%
  \BibitemOpen
  \bibfield  {author} {\bibinfo {author} {\bibfnamefont {F.~C.}\ \bibnamefont
  {Hoppensteadt}}\ and\ \bibinfo {author} {\bibfnamefont {E.~M.}\ \bibnamefont
  {Izhikevich}},\ }\href@noop {} {\emph {\bibinfo {title} {Weakly Connected
  Neural Networks}}}\ (\bibinfo  {publisher} {Springer},\ \bibinfo {address}
  {New York},\ \bibinfo {year} {1980})\BibitemShut {NoStop}%
\bibitem [{\citenamefont {Pikovsky}\ \emph {et~al.}(2001)\citenamefont
  {Pikovsky}, \citenamefont {Rosenblum},\ and\ \citenamefont
  {Kurths}}]{pikovsky2001}%
  \BibitemOpen
  \bibfield  {author} {\bibinfo {author} {\bibfnamefont {A.}~\bibnamefont
  {Pikovsky}}, \bibinfo {author} {\bibfnamefont {M.}~\bibnamefont {Rosenblum}},
  \ and\ \bibinfo {author} {\bibfnamefont {J.}~\bibnamefont {Kurths}},\ }\href
  {\doibase 10.1017/CBO9780511755743} {\emph {\bibinfo {title}
  {Synchronization: {A} {Universal} {Concept} in {Nonlinear} {Sciences}}}},\
  \bibinfo {edition} {1st}\ ed.\ (\bibinfo  {publisher} {Cambridge University
  Press},\ \bibinfo {year} {2001})\BibitemShut {NoStop}%
\bibitem [{\citenamefont {Ermentrout}\ and\ \citenamefont
  {Terman}(2010)}]{Ermentrout-Terman-10}%
  \BibitemOpen
  \bibfield  {author} {\bibinfo {author} {\bibfnamefont {G.~B.}\ \bibnamefont
  {Ermentrout}}\ and\ \bibinfo {author} {\bibfnamefont {D.~H.}\ \bibnamefont
  {Terman}},\ }\href@noop {} {\emph {\bibinfo {title} {Mathematical Foundations
  of Neuroscience}}}\ (\bibinfo  {publisher} {Springer},\ \bibinfo {address}
  {New York},\ \bibinfo {year} {2010})\BibitemShut {NoStop}%
\bibitem [{\citenamefont {Nakao}(2016)}]{nakao2016phase}%
  \BibitemOpen
  \bibfield  {author} {\bibinfo {author} {\bibfnamefont {H.}~\bibnamefont
  {Nakao}},\ }\href@noop {} {\bibfield  {journal} {\bibinfo  {journal}
  {Contemporary Physics}\ }\textbf {\bibinfo {volume} {57}},\ \bibinfo {pages}
  {188} (\bibinfo {year} {2016})}\BibitemShut {NoStop}%
\bibitem [{\citenamefont {Monga}\ \emph {et~al.}(2018)\citenamefont {Monga},
  \citenamefont {Wilson}, \citenamefont {Matchen},\ and\ \citenamefont
  {Moehlis}}]{Monga_Wilson-Matchen-Moehlis-18}%
  \BibitemOpen
  \bibfield  {author} {\bibinfo {author} {\bibfnamefont {B.}~\bibnamefont
  {Monga}}, \bibinfo {author} {\bibfnamefont {D.}~\bibnamefont {Wilson}},
  \bibinfo {author} {\bibfnamefont {T.}~\bibnamefont {Matchen}}, \ and\
  \bibinfo {author} {\bibfnamefont {J.}~\bibnamefont {Moehlis}},\ }\href@noop
  {} {\bibfield  {journal} {\bibinfo  {journal} {Biological Cybernetics}\ }
  (\bibinfo {year} {2018})}\BibitemShut {NoStop}%
\bibitem [{\citenamefont {Pietras}\ and\ \citenamefont
  {Daffertshofer}(2019)}]{pietras2019network}%
  \BibitemOpen
  \bibfield  {author} {\bibinfo {author} {\bibfnamefont {B.}~\bibnamefont
  {Pietras}}\ and\ \bibinfo {author} {\bibfnamefont {A.}~\bibnamefont
  {Daffertshofer}},\ }\href@noop {} {\bibfield  {journal} {\bibinfo  {journal}
  {Physics Reports}\ }\textbf {\bibinfo {volume} {819}},\ \bibinfo {pages} {1}
  (\bibinfo {year} {2019})}\BibitemShut {NoStop}%
\bibitem [{\citenamefont {Kurebayashi}\ \emph {et~al.}(2013)\citenamefont
  {Kurebayashi}, \citenamefont {Shirasaka},\ and\ \citenamefont
  {Nakao}}]{kurebayashi2013}%
  \BibitemOpen
  \bibfield  {author} {\bibinfo {author} {\bibfnamefont {W.}~\bibnamefont
  {Kurebayashi}}, \bibinfo {author} {\bibfnamefont {S.}~\bibnamefont
  {Shirasaka}}, \ and\ \bibinfo {author} {\bibfnamefont {H.}~\bibnamefont
  {Nakao}},\ }\href {\doibase 10.1103/PhysRevLett.111.214101} {\bibfield
  {journal} {\bibinfo  {journal} {Physical Review Letters}\ }\textbf {\bibinfo
  {volume} {111}},\ \bibinfo {pages} {214101} (\bibinfo {year}
  {2013})}\BibitemShut {NoStop}%
\bibitem [{\citenamefont {Wilson}\ and\ \citenamefont
  {Ermentrout}(2018{\natexlab{a}})}]{Wilson-Ermentrout-18}%
  \BibitemOpen
  \bibfield  {author} {\bibinfo {author} {\bibfnamefont {D.}~\bibnamefont
  {Wilson}}\ and\ \bibinfo {author} {\bibfnamefont {B.}~\bibnamefont
  {Ermentrout}},\ }\href@noop {} {\bibfield  {journal} {\bibinfo  {journal}
  {Biological Cybernetics}\ }\textbf {\bibinfo {volume} {76}},\ \bibinfo
  {pages} {37} (\bibinfo {year} {2018}{\natexlab{a}})}\BibitemShut {NoStop}%
\bibitem [{\citenamefont {Mauroy}\ and\ \citenamefont
  {Mez\'ic}(2018)}]{Mauroy-Mezic-18}%
  \BibitemOpen
  \bibfield  {author} {\bibinfo {author} {\bibfnamefont {A.}~\bibnamefont
  {Mauroy}}\ and\ \bibinfo {author} {\bibfnamefont {I.}~\bibnamefont
  {Mez\'ic}},\ }\href {\doibase 10.1063/1.5030175} {\bibfield  {journal}
  {\bibinfo  {journal} {Chaos}\ }\textbf {\bibinfo {volume} {28}},\ \bibinfo
  {pages} {073108} (\bibinfo {year} {2018})}\BibitemShut {NoStop}%
\bibitem [{\citenamefont {Wilson}\ and\ \citenamefont
  {Ermentrout}(2019)}]{wilson2019c}%
  \BibitemOpen
  \bibfield  {author} {\bibinfo {author} {\bibfnamefont {D.}~\bibnamefont
  {Wilson}}\ and\ \bibinfo {author} {\bibfnamefont {B.}~\bibnamefont
  {Ermentrout}},\ }\href {\doibase 10.1103/PhysRevLett.123.164101} {\bibfield
  {journal} {\bibinfo  {journal} {Physical Review Letters}\ }\textbf {\bibinfo
  {volume} {123}},\ \bibinfo {pages} {164101} (\bibinfo {year}
  {2019})}\BibitemShut {NoStop}%
\bibitem [{\citenamefont {Rosenblum}\ and\ \citenamefont
  {Pikovsky}(2019{\natexlab{a}})}]{rosenblum2019}%
  \BibitemOpen
  \bibfield  {author} {\bibinfo {author} {\bibfnamefont {M.}~\bibnamefont
  {Rosenblum}}\ and\ \bibinfo {author} {\bibfnamefont {A.}~\bibnamefont
  {Pikovsky}},\ }\href {\doibase 10.1063/1.5079617} {\bibfield  {journal}
  {\bibinfo  {journal} {Chaos: An Interdisciplinary Journal of Nonlinear
  Science}\ }\textbf {\bibinfo {volume} {29}},\ \bibinfo {pages} {011105}
  (\bibinfo {year} {2019}{\natexlab{a}})}\BibitemShut {NoStop}%
\bibitem [{\citenamefont {Rosenblum}\ and\ \citenamefont
  {Pikovsky}(2019{\natexlab{b}})}]{Rosenblum-Pikovsky-19a}%
  \BibitemOpen
  \bibfield  {author} {\bibinfo {author} {\bibfnamefont {M.}~\bibnamefont
  {Rosenblum}}\ and\ \bibinfo {author} {\bibfnamefont {A.}~\bibnamefont
  {Pikovsky}},\ }\href@noop {} {\bibfield  {journal} {\bibinfo  {journal}
  {Phil. Trans. R. Soc. A}\ }\textbf {\bibinfo {volume} {377}},\ \bibinfo
  {pages} {20190093} (\bibinfo {year} {2019}{\natexlab{b}})}\BibitemShut
  {NoStop}%
\bibitem [{\citenamefont {Le\'on}\ and\ \citenamefont
  {Paz\'o}(2019)}]{Leon-Pazo-19}%
  \BibitemOpen
  \bibfield  {author} {\bibinfo {author} {\bibfnamefont {I.}~\bibnamefont
  {Le\'on}}\ and\ \bibinfo {author} {\bibfnamefont {D.}~\bibnamefont
  {Paz\'o}},\ }\href@noop {} {\bibfield  {journal} {\bibinfo  {journal} {Phys.
  Rev. E}\ }\textbf {\bibinfo {volume} {100}},\ \bibinfo {pages} {012211}
  (\bibinfo {year} {2019})}\BibitemShut {NoStop}%
\bibitem [{\citenamefont {P\'erez-{C}ervera}\ \emph {et~al.}(2020)\citenamefont
  {P\'erez-{C}ervera}, \citenamefont {{M}-{S}eara},\ and\ \citenamefont
  {Huguet}}]{Perez-Seara-Huguet-20}%
  \BibitemOpen
  \bibfield  {author} {\bibinfo {author} {\bibfnamefont {A.}~\bibnamefont
  {P\'erez-{C}ervera}}, \bibinfo {author} {\bibfnamefont {T.}~\bibnamefont
  {{M}-{S}eara}}, \ and\ \bibinfo {author} {\bibfnamefont {G.}~\bibnamefont
  {Huguet}},\ }\href {\doibase 10.1063/5.0010149} {\bibfield  {journal}
  {\bibinfo  {journal} {Chaos}\ }\textbf {\bibinfo {volume} {30}},\ \bibinfo
  {pages} {083117} (\bibinfo {year} {2020})}\BibitemShut {NoStop}%
\bibitem [{\citenamefont {Gengel}\ \emph {et~al.}(2021)\citenamefont {Gengel},
  \citenamefont {Teichmann}, \citenamefont {Rosenblum},\ and\ \citenamefont
  {Pikovsky}}]{gengel2021}%
  \BibitemOpen
  \bibfield  {author} {\bibinfo {author} {\bibfnamefont {E.}~\bibnamefont
  {Gengel}}, \bibinfo {author} {\bibfnamefont {E.}~\bibnamefont {Teichmann}},
  \bibinfo {author} {\bibfnamefont {M.}~\bibnamefont {Rosenblum}}, \ and\
  \bibinfo {author} {\bibfnamefont {A.}~\bibnamefont {Pikovsky}},\ }\href
  {\doibase 10.1088/2632-072X/abbed2} {\bibfield  {journal} {\bibinfo
  {journal} {Journal of Physics: Complexity}\ }\textbf {\bibinfo {volume}
  {2}},\ \bibinfo {pages} {015005} (\bibinfo {year} {2021})}\BibitemShut
  {NoStop}%
\bibitem [{\citenamefont {Kurebayashi}\ \emph {et~al.}(2022)\citenamefont
  {Kurebayashi}, \citenamefont {Yamamoto}, \citenamefont {Shirasaka},\ and\
  \citenamefont {Nakao}}]{kurebayashi2022}%
  \BibitemOpen
  \bibfield  {author} {\bibinfo {author} {\bibfnamefont {W.}~\bibnamefont
  {Kurebayashi}}, \bibinfo {author} {\bibfnamefont {T.}~\bibnamefont
  {Yamamoto}}, \bibinfo {author} {\bibfnamefont {S.}~\bibnamefont {Shirasaka}},
  \ and\ \bibinfo {author} {\bibfnamefont {H.}~\bibnamefont {Nakao}},\ }\href
  {\doibase 10.1103/PhysRevResearch.4.043176} {\bibfield  {journal} {\bibinfo
  {journal} {Physical Review Research}\ }\textbf {\bibinfo {volume} {4}},\
  \bibinfo {pages} {043176} (\bibinfo {year} {2022})}\BibitemShut {NoStop}%
\bibitem [{\citenamefont {Bick}\ \emph {et~al.}(2023)\citenamefont {Bick},
  \citenamefont {Böhle},\ and\ \citenamefont {Kuehn}}]{bick2023higherorder}%
  \BibitemOpen
  \bibfield  {author} {\bibinfo {author} {\bibfnamefont {C.}~\bibnamefont
  {Bick}}, \bibinfo {author} {\bibfnamefont {T.}~\bibnamefont {Böhle}}, \ and\
  \bibinfo {author} {\bibfnamefont {C.}~\bibnamefont {Kuehn}},\ }\href@noop {}
  {\enquote {\bibinfo {title} {Higher-order interactions in phase oscillator
  networks through phase reductions of oscillators with phase dependent
  amplitude},}\ } (\bibinfo {year} {2023}),\ \Eprint
  {http://arxiv.org/abs/2305.04277} {arXiv:2305.04277 [math.DS]} \BibitemShut
  {NoStop}%
\bibitem [{\citenamefont {Shilnikov}\ \emph {et~al.}(1998)\citenamefont
  {Shilnikov}, \citenamefont {Shilnikov}, \citenamefont {Turaev},\ and\
  \citenamefont {Chua}}]{Shilnikov_Shilnikov_Turaev_Chua-98}%
  \BibitemOpen
  \bibfield  {author} {\bibinfo {author} {\bibfnamefont {L.}~\bibnamefont
  {Shilnikov}}, \bibinfo {author} {\bibfnamefont {A.}~\bibnamefont
  {Shilnikov}}, \bibinfo {author} {\bibfnamefont {D.}~\bibnamefont {Turaev}}, \
  and\ \bibinfo {author} {\bibfnamefont {L.}~\bibnamefont {Chua}},\ }\href@noop
  {} {\emph {\bibinfo {title} {Methods of Qualitative Theory in Nonlinear
  Dynamics ({P}art I)}}}\ (\bibinfo  {publisher} {World Scientific},\ \bibinfo
  {address} {Singapure},\ \bibinfo {year} {1998})\BibitemShut {NoStop}%
\bibitem [{\citenamefont {Wilson}\ and\ \citenamefont
  {Ermentrout}(2018{\natexlab{b}})}]{wilson2018a}%
  \BibitemOpen
  \bibfield  {author} {\bibinfo {author} {\bibfnamefont {D.}~\bibnamefont
  {Wilson}}\ and\ \bibinfo {author} {\bibfnamefont {B.}~\bibnamefont
  {Ermentrout}},\ }\href {\doibase 10.1007/s00285-017-1141-6} {\bibfield
  {journal} {\bibinfo  {journal} {Journal of Mathematical Biology}\ }\textbf
  {\bibinfo {volume} {76}},\ \bibinfo {pages} {37} (\bibinfo {year}
  {2018}{\natexlab{b}})}\BibitemShut {NoStop}%
\bibitem [{\citenamefont {Turaev}(2023)}]{DT}%
  \BibitemOpen
  \bibfield  {author} {\bibinfo {author} {\bibfnamefont {D.}~\bibnamefont
  {Turaev}},\ }\href@noop {} {\enquote {\bibinfo {title} {Private
  communication},}\ } (\bibinfo {year} {2023})\BibitemShut {NoStop}%
\bibitem [{\citenamefont {Ashwin}\ and\ \citenamefont
  {Rodrigues}(2016)}]{ashwin2016hopf}%
  \BibitemOpen
  \bibfield  {author} {\bibinfo {author} {\bibfnamefont {P.}~\bibnamefont
  {Ashwin}}\ and\ \bibinfo {author} {\bibfnamefont {A.}~\bibnamefont
  {Rodrigues}},\ }\href@noop {} {\bibfield  {journal} {\bibinfo  {journal}
  {Physica D: Nonlinear Phenomena}\ }\textbf {\bibinfo {volume} {325}},\
  \bibinfo {pages} {14} (\bibinfo {year} {2016})}\BibitemShut {NoStop}%
\bibitem [{\citenamefont {Nijholt}\ \emph {et~al.}(2022)\citenamefont
  {Nijholt}, \citenamefont {Ocampo-Espindola}, \citenamefont {Eroglu},
  \citenamefont {Kiss},\ and\ \citenamefont {Pereira}}]{nijholt2022emergent}%
  \BibitemOpen
  \bibfield  {author} {\bibinfo {author} {\bibfnamefont {E.}~\bibnamefont
  {Nijholt}}, \bibinfo {author} {\bibfnamefont {J.~L.}\ \bibnamefont
  {Ocampo-Espindola}}, \bibinfo {author} {\bibfnamefont {D.}~\bibnamefont
  {Eroglu}}, \bibinfo {author} {\bibfnamefont {I.~Z.}\ \bibnamefont {Kiss}}, \
  and\ \bibinfo {author} {\bibfnamefont {T.}~\bibnamefont {Pereira}},\
  }\href@noop {} {\bibfield  {journal} {\bibinfo  {journal} {Nature
  communications}\ }\textbf {\bibinfo {volume} {13}},\ \bibinfo {pages} {4849}
  (\bibinfo {year} {2022})}\BibitemShut {NoStop}%
\bibitem [{\citenamefont {Wilson}(2020)}]{wilson2020a}%
  \BibitemOpen
  \bibfield  {author} {\bibinfo {author} {\bibfnamefont {D.}~\bibnamefont
  {Wilson}},\ }\href {\doibase 10.1103/PhysRevE.101.022220} {\bibfield
  {journal} {\bibinfo  {journal} {Physical Review E}\ }\textbf {\bibinfo
  {volume} {101}},\ \bibinfo {pages} {022220} (\bibinfo {year}
  {2020})}\BibitemShut {NoStop}%
\bibitem [{Note1()}]{Note1}%
  \BibitemOpen
  \bibinfo {note} {We fix $\tau = 2000$ and $t_0 = 2000$. The initial states
  for the phase models are $(\phi _1, \phi _2) = (0,0)$ and the initial state
  for the full model is $(x,y,\phi _2) = (1,0,0)$.}\BibitemShut {Stop}%
\bibitem [{\citenamefont {Kumar}\ and\ \citenamefont
  {Rosenblum}(2021)}]{PhysRevE.104.054202}%
  \BibitemOpen
  \bibfield  {author} {\bibinfo {author} {\bibfnamefont {M.}~\bibnamefont
  {Kumar}}\ and\ \bibinfo {author} {\bibfnamefont {M.}~\bibnamefont
  {Rosenblum}},\ }\href {\doibase 10.1103/PhysRevE.104.054202} {\bibfield
  {journal} {\bibinfo  {journal} {Phys. Rev. E}\ }\textbf {\bibinfo {volume}
  {104}},\ \bibinfo {pages} {054202} (\bibinfo {year} {2021})}\BibitemShut
  {NoStop}%
\bibitem [{\citenamefont {von~der Gracht}\ \emph {et~al.}(2023)\citenamefont
  {von~der Gracht}, \citenamefont {Nijholt},\ and\ \citenamefont
  {Rink}}]{vondergracht2023parametrisation}%
  \BibitemOpen
  \bibfield  {author} {\bibinfo {author} {\bibfnamefont {S.}~\bibnamefont
  {von~der Gracht}}, \bibinfo {author} {\bibfnamefont {E.}~\bibnamefont
  {Nijholt}}, \ and\ \bibinfo {author} {\bibfnamefont {B.}~\bibnamefont
  {Rink}},\ }\href@noop {} {\enquote {\bibinfo {title} {A parametrisation
  method for high-order phase reduction in coupled oscillator networks},}\ }
  (\bibinfo {year} {2023}),\ \Eprint {http://arxiv.org/abs/2306.03320}
  {arXiv:2306.03320 [math.DS]} \BibitemShut {NoStop}%
\end{thebibliography}
\end{document}